\documentclass[12pt,twocolumn]{aastex63}

\usepackage{amsmath}

\shorttitle{NARROWBAND SPIKES} \shortauthors{Karlick\'y \& Ben\'a\v{c}ek}
\begin{document}

\title{Narrowband Spikes Observed during the 2013 November 7 Flare}

\author[0000-0002-3963-8701]{Marian Karlick\'y}
\affil{Astronomical Institute of the Academy of Sciences of the Czech Republic,
Fri\v{c}ova 298, CZ-25165 Ond\v{r}ejov, Czech Republic}
\email{karlicky@asu.cas.cz}

\author[0000-0002-4319-8083]{Jan Ben\'a\v{c}ek}
\affil{Center for Astronomy and Astrophysics, Technical University of Berlin, 10623 Berlin,
Germany}
\email{benacek@tu-berlin.de}

\author[0000-0003-3128-8396]{J\'an Ryb\'ak}
\affil{Astronomical Institute, Slovak Academy of Sciences,
  SK-05960 Tatransk\'{a} Lomnica, Slovakia}
\email{rybak@astro.sk}

\begin{abstract}
Narrowband spikes are observed in solar flares for several decades.
However, their exact origin is still discussed. To contribute to understanding
of these spikes, we analyze the narrowband spikes observed in the 800-2000 MHz
range during the impulsive phase of the November 7, 2013 flare. In the radio
spectrum, the spikes started with typical broadband clouds of spikes, and then
their distribution in frequencies changed into unique, very narrow bands having
non-integer frequency ratios. We successfully fitted frequencies of these
narrow spike bands by those, calculating dispersion branches and growth rates
of the Bernstein modes. For comparison, we also analyzed the model, where the
narrow bands of spikes are generated at the upper-hybrid frequencies. Using
both models, we estimated the plasma density and magnetic field in spike
sources. Then the models are discussed, and arguments in favor of the model
with the Bernstein modes are presented. Analyzing frequency profiles of this
spike event by the Fourier method, we found the power-law spectra with the
power-law indices varying in the -0.8 -- -2.75 interval. Because at some times
this power-law index was close to the Kolmogorov spectral index (-5/3), we
propose that the spikes are generated through the Bernstein modes in turbulent
plasma reconnection outflows or directly in the turbulent magnetic reconnection
of solar flares.
\end{abstract}


\section{Introduction}

Radio bursts are an integral part of solar flares.  Their types and basic
characteristics are well described, e.g., in books by
\cite{1985srph.book.....M,1979itsr.book.....K} and papers by
\cite{1994A&AS..104..145I} and \cite{2001A&A...375..243J}. Among these bursts
the narrowband dm-spikes belong to those of the most interesting, as they seem
to be connected with the primary flare energy-release processes
\citep{1977A&A....57..285D,1984SoPh...92..329K,1985spit.conf..560F,1986SoPh..104..117S,1990A&A...239L...1G,
1994A&A...285.1038K,1998ApJ...497..972A,1998SoPh..182..477Z,2016A&A...586A..29B}.
On the radio spectrum, spikes occur in clouds of many short duration narrowband
bursts with a typical relative bandwidth of about 1-3 $\%$, duration less than
100 ms and the brightness temperature up to 10$^{15}$ K
\citep{1986SoPh..104...99B}. In some cases, the narrowband spikes are observed
near the starting frequency of type III bursts and in a good correlation with
the hard X-ray emission \citep{2009A&A...504..565D}.

Several models of the narrowband spikes were proposed. For example, in papers
by \citet{1981A&A...103..331K,1990ApJ...353..666T,1991ApJ...373..285W} the
runaway electrons, accelerated in a strong DC electric field, are considered as
the primary source of the spikes. In other models, the electron-cyclotron maser
(ECM) mechanism in connection with the loss-cone distribution of superthermal
electrons was suggested. In papers by \cite{1982ApJ...259..844M},
\cite{1985ApJ...290..347V}, \cite{1988ApJ...328..809W},
\cite{1990A&AS...85.1141A}, \cite{1998PhyU...41.1157F},
\cite{2003ApJ...593..571F} and \cite{2017RvMPP...1....5M} assuming the ratio $Y
= \omega_\mathrm{pe}/\omega_\mathrm{ce} < 1$, where $\omega_\mathrm{pe}$ and
$\omega_\mathrm{ce}$ are the electron plasma and electron cyclotron
frequencies, the electromagnetic waves (spikes) are generated directly by this
ECM mechanism. In first versions of this mechanism the emission occurs near the
cyclotron frequency and its harmonics. However, \cite{1994A&A...285.1038K}
presented observations, where the narrowband spikes were clustered in bands
with the non-integer ratio in the interval 1.06 - 1.54. Therefore,
\cite{1996ApJ...467..465W} proposed the model, where the spike frequencies
correspond to the Bernstein modes (BM model). Similarly,
\cite{1999ApJ...524..961S} and \cite{2001A&A...379.1045B} presented the model
of spikes with the upper-hybrid waves (UHW model). In these models the ratio $Y
$ is greater than unity. We note that in some conditions also the direct ECM
mechanism may generate non-integer harmonic ratios \citep{1998PhyU...41.1157F}.
Recently, using the particle-in-cell model, the plasma emission induced by the
ECM instability in solar plasmas with the ratio $Y$ = 10 have been studied by
\cite{2020ApJ...891L..25N}. They found that the fundamental emission is caused
by coalescence of almost counterpropagating Z-modes and whistlers, while the
harmonic emission arises from coalescence of an almost counterpropagating UH
waves.

For understanding of these spikes also the bandwidth and duration of
individual spikes were studied in details. For example, in the paper by
\cite{2008ApJ...681.1688R} the characteristic spike half-maximum duration
$\tau$ in dependence of the frequency $f$ was derived as $\tau \sim f^{-1.29}$.
On the other hand, the power-law distributions of spike bandwidths in
dependence on frequency were found by \cite{2014ApJ...789..152N}.

In papers by \cite{1996SoPh..168..375K,2000SoPh..195..165K}, the frequency
bandwidths of spikes were studied by the Fourier method. As a result the
power-law spectra with power-law indices in the range of -0.80 -- -2.85 were
found. Similar results were presented by \cite{2000A&A...354..287M}. Because in
several events the power-law indices were close to -5/3, it was suggested that
the narrowband dm-spikes are generated by superthermal electrons in the
magnetohydrodynamic (MHD) turbulence in the magnetic reconnection outflows
\citep{1996SoPh..168..375K,1998SoPh..182..477Z}.

Magnetic reconnection converts magnetic energy to other forms of energy,
driving thus solar flares and accelerating particles
\citep{2000mare.book.....P}. Its 2-D numerical models have shown that a current
layer during the reconnection becomes fragmented to smaller and smaller
magnetic islands (plasmoids)
\citep{2007PhPl...14j0703L,2009PhPl...16k2102B,2011ApJ...730...47B,2011ApJ...737...24B,2011ApJ...733..107K}.
On the other hand, 3-D numerical simulations \citep{2019PhPl...26a2901D} show
evolution of turbulence in agreement with the turbulent reconnection theory
\citep{1999ApJ...517..700L,2020PhPl...27a2305L}. Thus, the turbulence can
take place directly in the reconnection layer as well as in the plasma
reconnection outflows. In turbulent reconnection the electrons are effectively
accelerated as described by \cite{2014PhPl...21i2304D},
\cite{2015ApJ...806..167G}, \cite{2016ApJ...827...94Z} and
\cite{2020PhPl...27h0501G}.

In the present paper, we analyze the narrowband spikes observed during
the impulsive phase of the November 7, 2013 flare. In the radio spectrum, these
spikes firstly appear in clouds of spikes as in typical spike events, but then
the frequency distribution of these spikes changed to unique, very narrow bands
with the non-integer frequency ratios. Just these very narrow bands of spikes
with the non-integer frequency ratios enable us to verify some their models.
Calculating dispersion branches and growth rates from analytical equations, we
show that these spikes can be explained by the model with the Bernstein modes
proposed by \cite{1996ApJ...467..465W}. Using this model, we determine the
plasma density and magnetic field strength in a source of the spikes. For
comparison, we also computed the growth rates of the upper-hybrid waves in the
models proposed by \cite{1999ApJ...524..961S} and \cite{2001A&A...379.1045B}.
Comparing the both models, we conclude  that the model with the Bernstein modes
is more probable. We note that the transition from typical broadband
clouds of spikes to very narrow spike bands, observed in this spike event,
shows that the results obtained in the analysis of these narrow bands have also
a general significance at least for some group of spike events. In this paper,
we also analyze frequency profiles of these spikes by the Fourier method. Based
on the computed power-law spectra, we suggest that the spikes are generated in
turbulent plasma reconnection outflows or directly in turbulent magnetic
reconnection of solar flares.

\section{Observations and their analysis}

\begin{figure}
\begin{center}
\includegraphics[width=12cm]{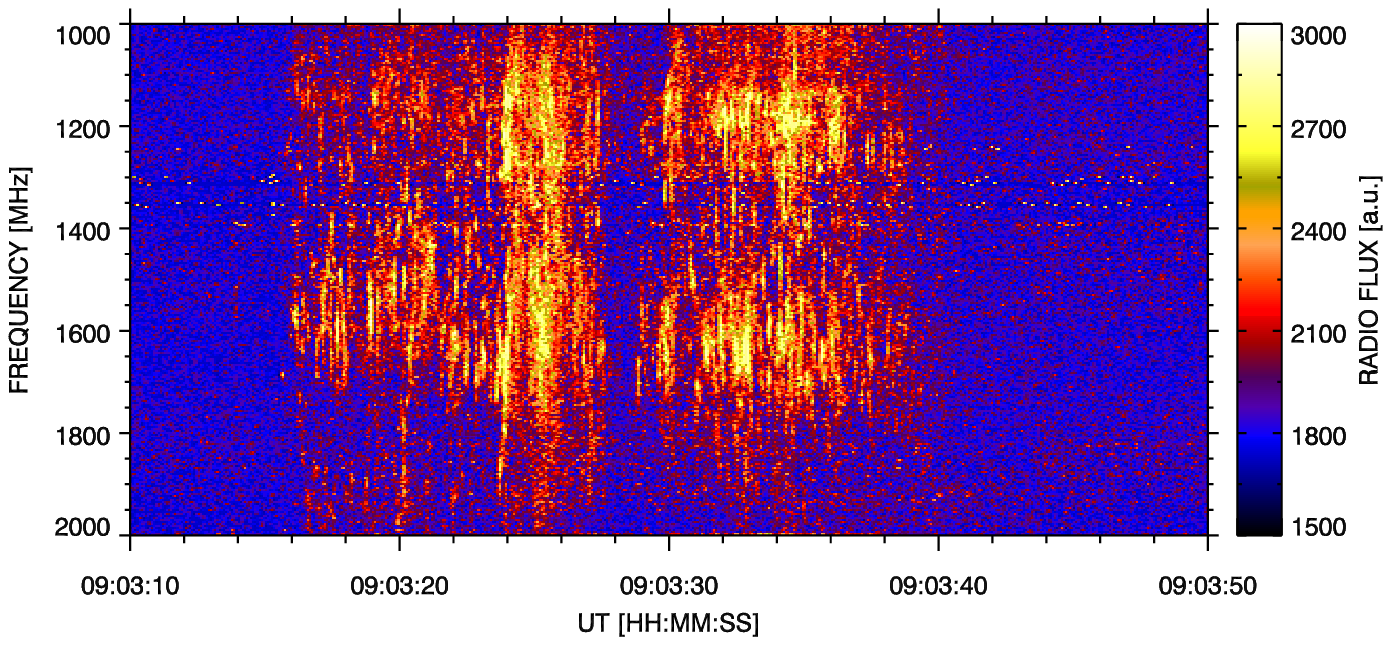}
\put(-295,35){\textcolor{white}{\bf a)}}

\includegraphics[width=12cm]{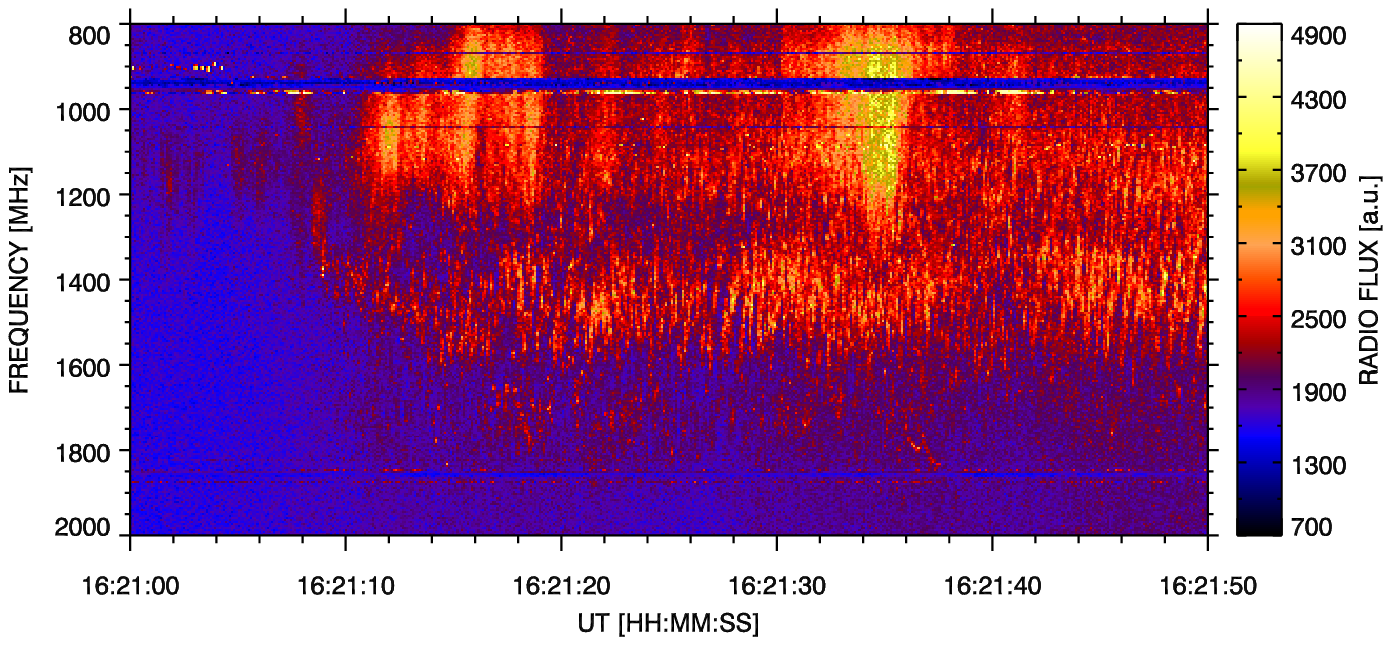}
\put(-295,35){\textcolor{white}{\bf b)}}
\end{center}
  \caption{Examples of the narrowband spikes observed in the 14 September 1992 (a) and 10 June 2003 (b) flares.
  The horizontal narrow band in the 940-970 MHz range of the 10 June 2003 spectrum means a gap in observations.}
  \label{fig1}
\end{figure}

During 1992-2020 years the Ond\v{r}ejov radiospectrographs
\citep{1993SoPh..147..203J,2008SoPh..253...95J}, operating in the 800-2000 MHz
range, registered 53 clouds of the narrowband spikes.  Two examples of the
narrowband spikes observed in the impulsive phase of the 14 September 1992 and
10 June 2003 flares are shown in Figure~\ref{fig1}. As can be seen here the
narrowband spikes are typically clustered in broad frequency bands; in these
cases at least in two bands. In the 10 June 2003 radio spectrum the narrowband
spikes are localized close to the starting frequency of type III bursts, see
Figure~\ref{fig1}b at frequencies below 1200 MHz. It indicates that the
narrowband spikes are generated close to the region where electrons are
accelerated.

\begin{figure}
\begin{center}
\includegraphics[width=12cm]{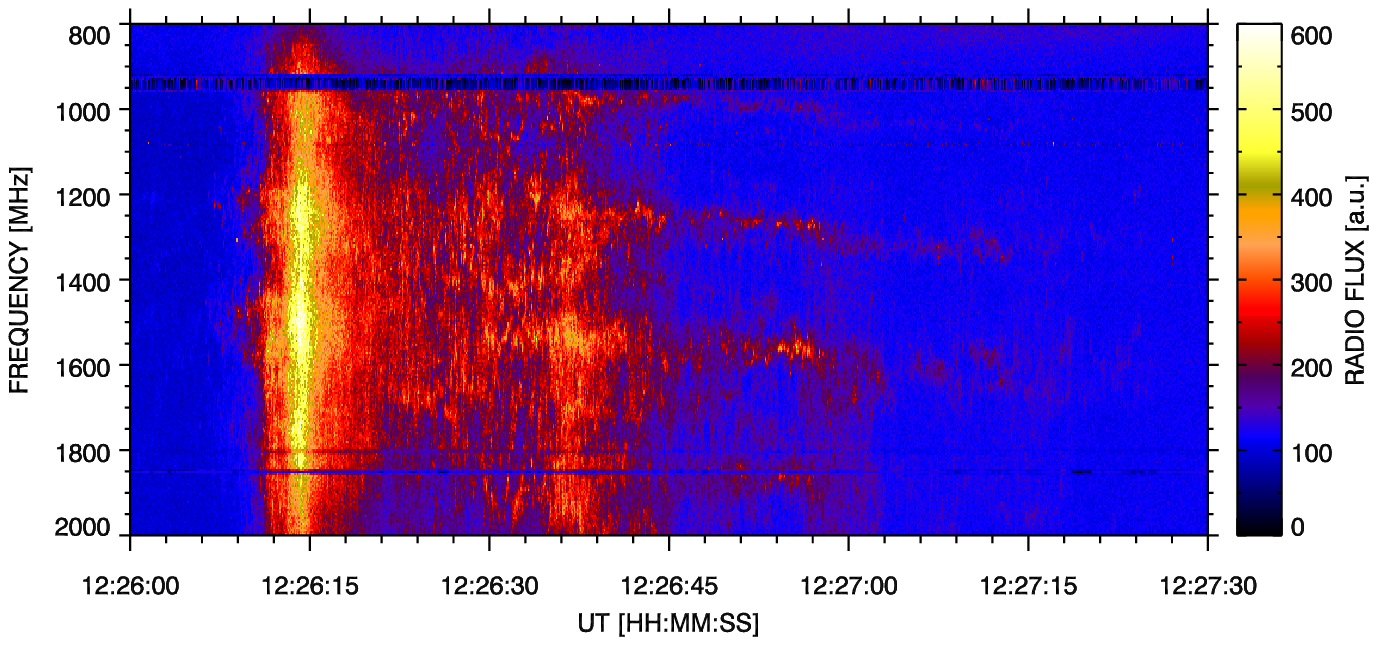}
\put(-295,35){\textcolor{white}{\bf a)}}

\includegraphics[width=12cm]{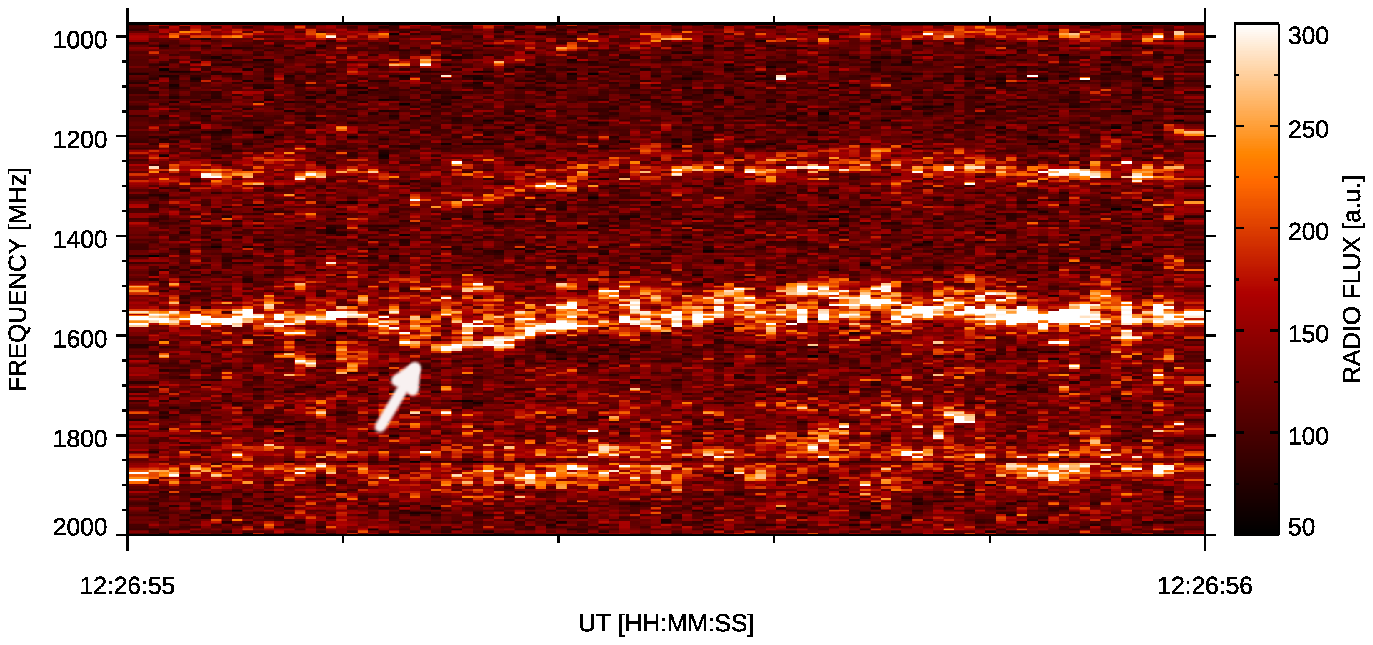}
\put(-295,35){\textcolor{white}{\bf b)}}

\includegraphics[width=12cm]{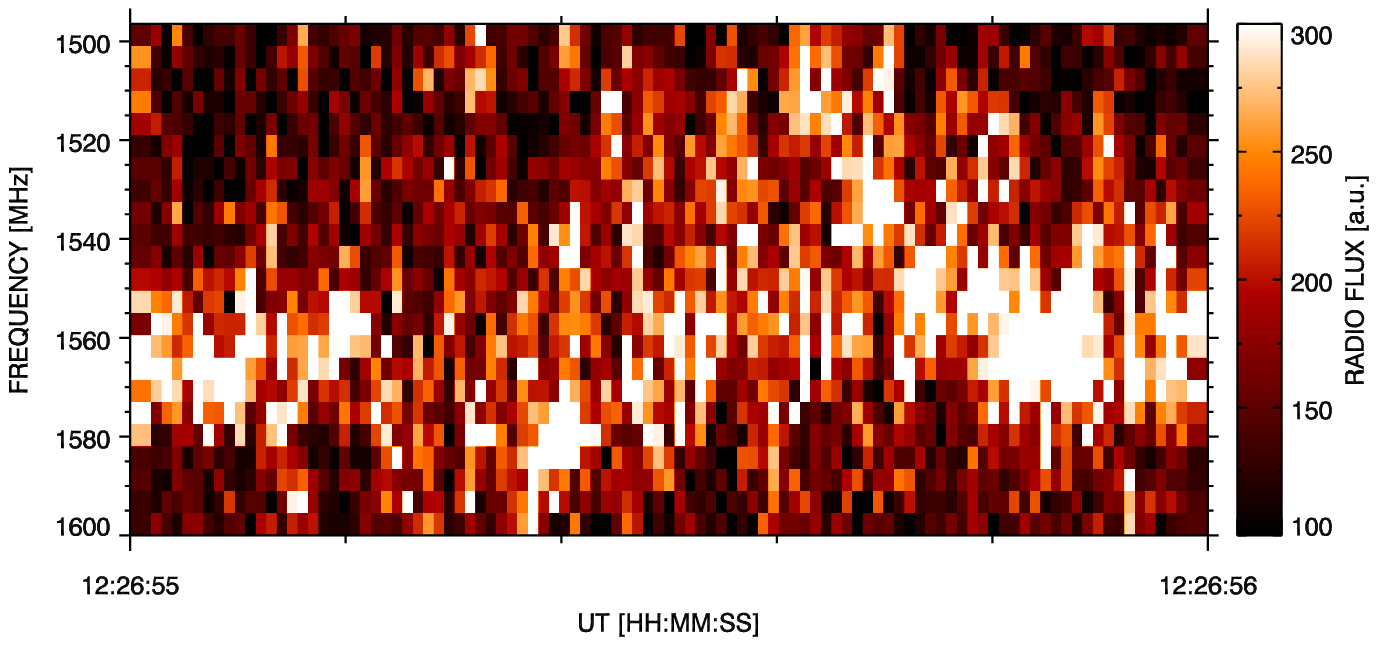}
\put(-295,35){\textcolor{white}{\bf c)}}
\end{center}
  \caption{a) Radio spectrum showing the narrowband spikes that started as typical spike event with spikes in broad bands (12:26:09 - 12:26:39 UT)
  and followed by unique, four very narrow bands of spikes observed on 7 November
  2013. In the 940-970 MHz range there is a gap in observations.
  b) Detail of this radio spectrum in the 1000-2000 MHz range starting at 12:26:55 UT, lasting 1 s and showing the narrow bands of spikes.
  The time resolution is 0.01 s. The arrow shows the band frequency variation
  which the form is synchronized with the similar forms in all four bands. The
  spectrum is not smoothed and thus showing bins in the record.
  c) The same as b), but in the 1500-1600 MHz range, showing bright bins (roughly spikes) within one narrow band of spikes.}
  \label{fig2}
\end{figure}

Among these 53 cases of the observed narrowband spikes, we found a
unique example of the narrowband spikes observed in November 7, 2013
(Figure~\ref{fig2}). The time and frequency resolution of the radio spectrum in
this case is 0.01 s and 4.7 MHz, respectively \citep{2008SoPh..253...95J}.
These spikes started at 12:26:09 UT with the broad bands of spikes. Around
12:26:14 UT these spikes are so densely distributed that nearly form a
continuum. Then the density of spikes in the radio spectrum decreases and
varies up to 12:26:39 UT. Whole this interval of spikes from 12:26:09 till
12:26:39 UT resembles to typical cases of the narrowband spikes in this
frequency range. After 12:26:39 UT four bands of spikes appeared and they were
very narrow. The spike event ends at about 12:27:20 UT. The part of the spike
event with very narrow bands of spikes is unique as well as the transition from
broadband bands of spikes to narrow ones. As will be seen in the following, the
part of the spike event with four very narrow bands of spikes enables
comparison with theoretical models of the narrowband spikes. On the other hand
the transition from broad bands of spikes to very narrow bands shows that the
results obtained for spikes in the narrow bands are also valid for spikes at
the beginning part of the spike event (at 12:26:09 till 12:26:39 UT) that
resembles to typical spike events.

These narrowband spikes presented in Figure~\ref{fig2}a were observed in the
impulsive phase of the November 7, 2013 flare that occurred in the NOAA AR 1890
in H$_{\alpha}$ with the start at 12:17 UT, maximum at 12:28 UT, and end at
12:44 UT, and in GOES soft X-ray emission with the start at 12:22 UT, maximum
at 12:29 UT, and end at 12:34 UT. At these times the Atmospheric Imaging
Assembly (AIA) \citep{2012SoPh..275...17L} onboard the {\it Solar Dynamic
Observatory} (SDO) \citep{2012SoPh..275....3P} shows a compact flare with many
interacting multi-thermal loops. There is a gap in RHESSI hard X-ray
observations.

Details of the very narrow bands of spikes with the maximal time resolution
(0.01 s) and in the time interval of one second starting at 12:26:55 UT are
shown in Figure~\ref{fig2}b and \ref{fig2}c. These spectra are without
any smoothing and thus showing bins with the duration 0.01 s and frequency
width 4.7 MHz. The characteristic half-maximum duration of spikes in the
frequency range under study is about 0.01 s
\citep{2003A&A...407.1115M,2008ApJ...681.1688R}. On the other hand, the mean
bandwidth of individual spikes was reported by \cite{2014ApJ...789..152N} as
7.5 MHz. Considering these values and time and frequency resolution of the
spectra, it means that bright bins roughly correspond to individual spikes. It
also means that these spikes are typical spikes. However, in this event they
are clustered in unique, very narrow bands, where individual spikes are
overlapping in some locations. The frequency of these narrow bands at 12:26:55
UT is about 1003, 1276, 1572 and 1877 MHz. It gives the ratio between
frequencies of the neighboring bands as 1877 MHz/1572 MHz = 1.19, 1572 MHz/1276
MHz = 1.23, and 1275 MHz/1000 MHz = 1.275. For comparison, the
frequency ratio between neighboring spike bands reported by
\cite{1994A&A...285.1038K} is in the 1.06-1.54 range. Thus, our values of this
ratio are within this range. Further interesting aspect of these narrow bands
can be seen in Figure~\ref{fig2}b. Here, in the time interval 12:26:55.2 -
12:26:55.4 UT there is the band frequency variation (see the arrow), form of
which is synchronized with the similar forms in all four bands. No further
narrow bands at this time and at frequencies below 800 MHz (Callisto BLEN
200-900 MHz spectrum\footnote{http://www.e-callisto.org/}) and above 2000 MHz
(Ond\v{r}ejov 2000-5000 MHz spectrum) were found.

\begin{figure}
\begin{center}
\includegraphics[width=16cm]{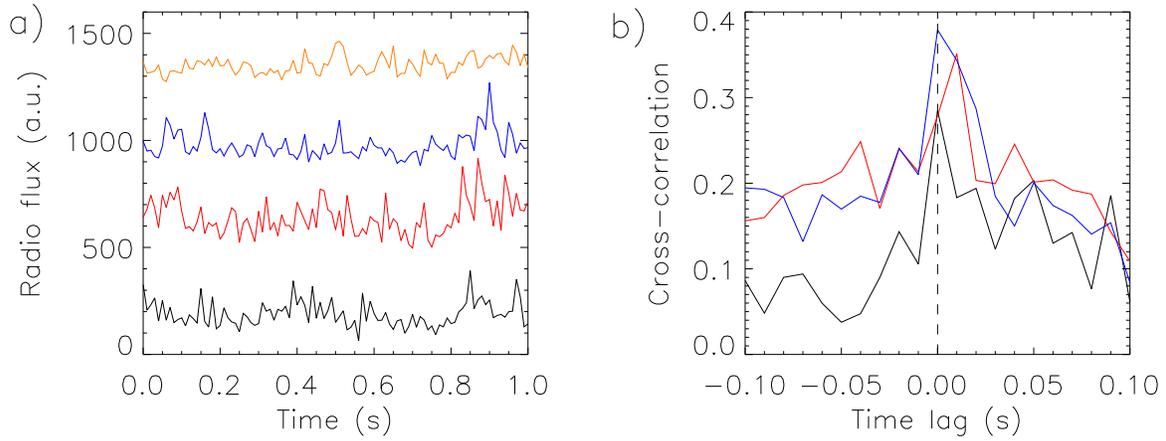}
\end{center}
  \caption{a) Radio flux in dependence on time at 1877 MHz (black line), at 1572 MHz
  (red line) + 400 a.u., at 1276 MHz (blue line) + 800 a.u., and
  at 1003 MHz (orange line) + 1200 a.u..
  starting at 12:26:55 UT and lasting 1 s. b) Cross-correlations of the radio flux profiles on 1877 and 1572 MHz (black line),
  1572 and 1276 MHz (red line), and 1276 and 1003 MHz (blue line).}
  \label{fig3}
\end{figure}

Figure~\ref{fig3}a shows the radio flux evolution in the time interval 1 s
starting at 12:26:55 UT at four frequencies (1003, 1276, 1572 and 1877 MHz)
that approximately cut the bands of spikes presented in Figure~\ref{fig2}b. In
all these radio flux records, there are variations on the shortest time scale
0.02 s (twice the temporal resolution). To know the relation between spikes in
these bands with the shortest time resolution (0.01 s), we calculated
cross-correlations between the radio flux profiles at these bands. The maximum
cross-correlations are shown in Figure~\ref{fig3}b: 0.38 between the radio
fluxes at frequencies 1276 MHz and 1003 MHz (blue line), 0.35 at frequencies
1572 MHz and 1276 MHz (red line), and 0.28 at frequencies  1877 MHz and 1572
MHz (black line). In two cross-correlations the time lag was zero (black and
blue line), and in the third case (red line) the time lag was 0.01 s (profile
at 1572 MHz was delayed 0.01 s after that on 1276 MHz). The cross-correlations
between channels from not neighboring spike bands were lower. The maximum
cross-correlations are low, caused by the low value of the signal/noise ratio
at this 0.01 s time resolution and overlapping of spikes, but with the
peaks clearly visible. Moreover, as mentioned above there are variations of
band frequencies that are synchronized in all four bands.

\begin{figure}
\begin{center}
\includegraphics[width=16cm]{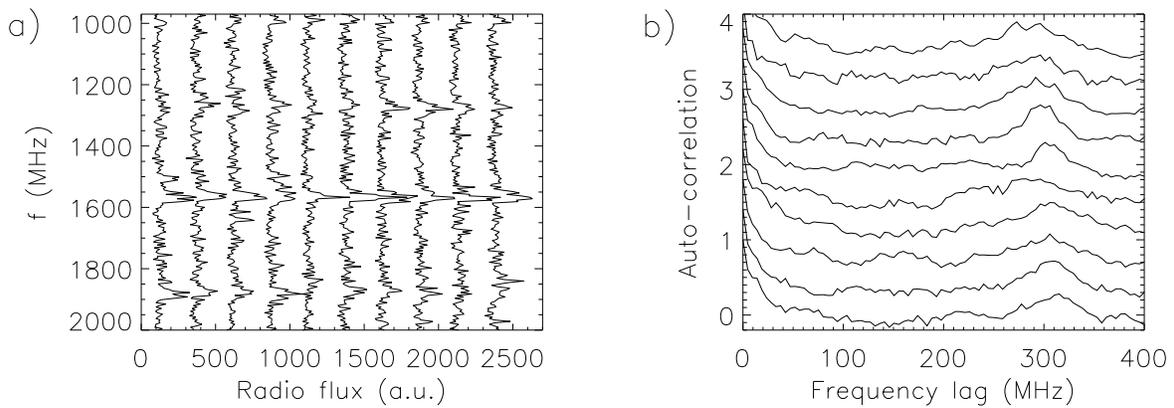}
\end{center}
  \caption{a) Radio flux in dependence on frequency in 10 instants starting at 12:26:55 UT, each after 0.01 s, corresponding to the radio flux shift
  250 a.u. b) Auto-correlations of the radio flux profiles shown in (a). The shift of each auto-correlation is
0.4.}
  \label{fig4}
\end{figure}

We also made 10 profiles of the radio flux vs. frequency for the spikes
observed during 12:26:55.00 - 12:26:55.09 UT (Figure~\ref{fig4}a). The time
difference between neighboring profiles is 0.01 s. Auto-correlations of these
profiles, showing the frequency lag of about 300 MHz, are shown in
Figure~\ref{fig4}b.

~

~

\section{Models}

Owing to a very narrow bandwidth of the four bands of spikes and
non-integer ratio of their neighboring frequencies, the 7 November 2013 event
is a good example for model verification. Because of this non-integer ratio and
because in the solar corona the mean ratio
$\omega_\mathrm{pe}/\omega_\mathrm{ce}$ is of the order of unity or larger (see
the models of the plasma density and magnetic field in the corona in the book
by \cite{2004psci.book.....A}), in the following, we analyze the models that
are relevant to these conditions. (However, it is not possible to exclude that
in some localized regions in the solar corona this ratio is lower than unity.)

First, we consider the model by \cite{1996ApJ...467..465W}, where the
observed band frequencies of spikes are the Bernstein mode frequencies (BM
model). We calculate dispersion branches of the Bernstein modes and their
corresponding growth rates in a similar way as in the paper by
\cite{2019ApJ...881...21B}. We assume a plasma that consists of a Maxwellian
cold and dense background plasma and rare hot superthermal electrons having the
loss-cone DGH distribution \citep{1965PhRvL..14..131D}. Such a plasma is
unstable as shown in Appendix~A, where the analytical relations for
computations of the electrostatic dispersion branches together with their
growth rates are described in detail. Using these relations, we tried to fit the
observed spike band frequencies shown in Figure~\ref{fig2}b (1003, 1276, 1572
and 1877 MHz) in such a way that the dispersion branches intersect the positive
growth rate regions just at the observed frequencies. Plasma parameters for
this fitting procedure were taken to be appropriate to solar flare conditions.
A good agreement between the observed and model frequencies was found for the
following parameters: $\omega_\mathrm{pe}/\omega_\mathrm{ce}=2.7,
v_\mathrm{tb}/c=0.02$ (2.38~MK), $v_\mathrm{t}/c=0.25$, and
$n_\mathrm{e}/n_\mathrm{h} = 10$, where $v_\mathrm{tb}$ is the thermal velocity
of the Maxwellian background plasma, $v_\mathrm{t}$ is the characteristic
velocity of the superthermal electrons, $c$ is the speed of light, and
$n_\mathrm{e}$ and $n_\mathrm{h}$ are the background cold plasma and hot plasma
densities.

The spikes in BM model can be generated on BM frequencies by the
process BM $\pm$ S $\rightarrow$ T, where BM is the Bernstein mode, S is the
low-frequency wave and T is the transversal electromagnetic (radio) wave, or on
the double BM frequencies by the process BM + BM $\rightarrow$ T.
\cite{1996ApJ...467..465W} preferred the emission on the double BM frequencies
saying that this process does not require additional low-frequency waves.
However, as shown by \cite{2020ApJ...891L..25N} the low-frequency waves
(whistlers and ion-acoustic) are present in such processes. Therefore in the
following estimations of the plasma density and magnetic field, we consider both
the possible processes.

The result of the above described fitting procedure with the assumption
that the radio emission frequencies of the spike bands correspond to BM
frequencies is shown in Figure~\ref{fig5}, i.e. the frequencies in MHz are
given for $\omega_\mathrm{pe}= 2 \pi f_\mathrm{pe}$, $f_\mathrm{pe} = 855$~MHz.
(We note that for the emission on the double BM frequencies the values of $f$
in MHz should be divided by 2 and $f_\mathrm{pe}$ = 427.5 MHz.) Positive
growth rates in this figure are expressed by the blue-red regions that are
approximately at harmonics of the cyclotron frequency. When the dispersion
branch (green line) intersects a positive growth rate region, the Bernstein
(electrostatic) modes are generated. In our case it is for the Bernstein modes
with the gyro-harmonic number $s$ = 3, 4, 5, and 6, corresponding to the spike
band frequencies 1003 MHz, 1276 MHz, 1572 MHz and 1877 MHz, respectively. While
the fit of three bands (1276, 1572 and 1877 MHz) is nearly exact, there is some
deviation between the observed band frequency at about 1003 MHz and that found
by fitting 950 MHz. But, note that this branch is close to the frequency gap in
observations and the parameters in computations can also deviate from real
ones.

Now taking the spike band frequencies as BM frequencies and using the
relations $f_\mathrm{pe} [\rm{MHz}] = 9 \times 10^{-3} \sqrt{\it{n}_e
[\rm{cm}^{-3}]}$ and $f_\mathrm{ce} [\rm{MHz}] = 2.8 \times$$B$ [G], where
$f_\mathrm{ce} = \omega_\mathrm{ce}/2 \pi$ \citep{2018ApJ...867...28K} we
estimated the mean plasma density and magnetic field strength in the spike
source in the November 7, 2013 event as $n_\mathrm{e}$ = 9 $\times$ 10$^{9}$
cm$^{-3}$ and $B$ = 113 G, respectively. On the other hand, for the radio
emission on the double BM frequencies the values of the plasma density and
magnetic field strength are 4 and 2 times lower, respectively.

Now for comparison, let us consider the model by \cite{1999ApJ...524..961S} and
\cite{2001A&A...379.1045B}. In this model the observed band frequencies of
spikes are frequencies of UHW. Using
the equations presented in Appendix~A and varying the ratio $\omega_\mathrm{UH}
/ \omega_\mathrm{ce}$, we computed the growth rates of the upper-hybrid waves
in the interval of the gyro-harmonic number $s=3-6$, see Figure~\ref{fig6}.
Other parameters were same as in the BM model. Each growth rate $\Gamma$
in this figure corresponds to the maximal growth rate on the upper-hybrid wave
branch. The maximal growth rates are found for the ratios $\omega_\mathrm{UH} /
\omega_\mathrm{ce} = 2.84, 3.80, 4.75$, and $5.71$. Although the parameters and
approximations in the present computations of the growth rates differ from
those used by \cite{1999ApJ...524..961S}, the results are similar.

Contrary to the BM model proposed by \cite{1996ApJ...467..465W}, where all
bands of spikes are generated in one source, in the UHW model, each spike
band is generated in different regions with different ratios of the plasma
density and the magnetic field. This model is similar to that proposed for
zebra patterns \citep{2013SoPh..284..579Z,2018ApJ...867...28K}. Thus, generally
spike bands could be used for determining the plasma density and magnetic field
strength in spike band sources. But, a problem is how to determine the
gyro-harmonic numbers of the spike bands in this case. If only as an example we
assume that the frequencies of observed spike bands $f_\mathrm{UH} = 1003,
1276, 1572$, and $1877$~MHz are the upper-hybrid frequencies for $s$ = 3, 4, 5
and 6, then the plasma density and magnetic field strength in four sources
corresponding to four bands in the 7 November 2013 event can be estimated as
$n_\mathrm{e} = 1.1 \times 10^{10}, 1.9 \times 10^{10}, 2.9 \times 10^{10}$,
and $4.2 \times 10^{10}$~cm$^{-3}$ and $B = 119, 114, 112$, and $111$~G,
respectively. If on the other hand, in agreement with zebra pattern
observations, we assume the inverse sequence of $s$ = 6, 5, 4, 3 for these
spike bands then the plasma density and magnetic field strength in these four
sources are $n_\mathrm{e} = 1.2 \times 10^{10}, 1.9 \times 10^{10}, 2.9 \times
10^{10}$, $3.9 \times 10^{10}$~cm$^{-3}$ and $B = 60, 91, 140, 223$~G,
respectively.  Here, we used the relations
$n_\mathrm{e}[\mathrm{cm}^{-3}]=f_{\mathrm{UH}}^2(1-1/s^2)/8.1 \times 10^{-5}$
and $B[\mathrm{G}]= f_{\mathrm{UH}}/(2.8s)$, where $f_{\mathrm{UH}}$ is in MHz
\citep{2020A&A...638A..22K}. As follows from these estimations and relations
the estimated plasma density is only partly dependent on the assumed
gyro-harmonic number $s$. However, the magnetic field strongly depends on $s$,
and for higher $s$ the magnetic field decreases. On the other hand, for
the radio emission on the double UHW frequencies the values of the plasma
density and magnetic field strength are 4 and 2 times lower, respectively.
Comparing the estimated plasma density and magnetic field in the BM and UHW models,
we can see that the plasma density and magnetic in the BM model is roughly the same
as in the band on 1003 MHz with $s$ = 3 in UHW models, other values of the
plasma density and magnetic field in the UHW model differ. But note that in BM
model there is one source and in the UHW model four different sources. Moreover, we
do not know the gyro-harmonic numbers of bands of spikes in the UHW model.

\section{Discussions}

A question arises, which model better agrees with observed spikes: the BM
model or the UHW model? The most important argument in favor of the BM model is that we
succeeded (almost perfectly) to fit BM frequencies with the observed band
structure. Moreover, the frequency difference among three bands was close to
300 MHz. It also speaks in favor of the BM model, because it is not very probable
that the UHW model with three sources at different locations give such a result.

There is further argument in favor of the BM model. Namely, we found zero or 0.01 s
time lag in the cross-correlations between two neighboring spike bands
(Figure~\ref{fig3}b). Moreover, we found the band frequency variations
synchronized in all four bands, see the arrow in Figure~\ref{fig2}b. When we
take now the UHW model, where band sources are at different locations, and taking
the frequencies of spike bands and using the density model of the solar
atmosphere by \cite{2002SSRv..101....1A}, the distance between sources of
neighboring spike bands can be estimated as about 2000 km. When we assume the
Alfv\'en waves for synchronization of the spikes in the bands, that are at the
distance 2000 km, within 0.01 s then the requested Alfv\'en speed is 200000 km
s$^{-1}$, which is unrealistic.

Now, let us look if some answers can be also found in an analysis of
the bandwidth of individual spikes or bandwidth of spike bands. Namely,
considering the BM model proposed by \cite{1996ApJ...467..465W}, we expect that
increasing the gyro-harmonic number of Bernstein modes the bandwidth of
individual spike as well as the bandwidth of spike bands increases roughly as
an increase of the gyro-harmonic numbers, i.e., when we multiply the lowest and
highest frequency of some individual spike or band of spikes by the
gyro-harmonic number s, then also the difference between the lowest and highest
frequencies (bandwidth) increases as multiplied by s. We cannot verify this effect on the
bandwidth of individual spikes here because the frequency
resolution of our observations is only 4.7 MHz and the typical bandwidth of
spikes in this range is $\sim$7.5 MHz \citep{2014ApJ...789..152N}. However,
\cite{2019Ap&SS.364....4F} presented chains of spikes, where the bandwidth of
the spikes on higher frequencies was about 2.1 times larger than that on lower
frequencies. But, note that such an increase of the bandwidth of individual
spikes in dependence on frequency can be also explained in models with direct
ECM emission \citep{2004ApJ...601..559F}.

Nevertheless, we can compare the bandwidth of the bands of spikes. Therefore in
Figure~\ref{fig7}, we show two contours of bands of narrowband spikes observed
at about 12:26:15 UT. At this time two bands were recognized in the 1370 - 1620
MHz and 1170 - 1370 MHz range (red contour and red dashed horizontal lines) and
1370 - 1700 MHz and 1370 - 1100 MHz (blue contour and red dashed horizontal
lines). Thus, the ratio of bandwidths of these two bands is 1.25 (250
MHz/200 MHz) and 1.22 (330 MHz/270 MHz) for red and blue bands, respectively.
Because the ratio 1.25 (red band contours) corresponds to the ratio of
gyro-harmonic numbers $s$=5 and $s$=4 (5/4 = 1.25) which were found for the
very narrow bands that followed these bands of spikes (Figure~\ref{fig2}a,b and
\ref{fig5}), this result supports the BM model. However, this result needs to
be taken with a caution because the bandwidth of spike bands can be influenced
by an overlapping of these bands and also by variations of the spike
intensities. We tried to check this result also in other spike events, but only
a few of them were without these overlapping and thus appropriate for such an
analysis. As a result, we found one example with the similar ratio as that
presented here, and one opposite case, where the spike band on higher frequency
was narrower than that on lower frequency, i.e., the ratio of these bands was
less than 1. Summarizing this part of discussions, we think that BM
model very well explains spikes in the November 7, 2013 event.

The present spikes were observed in the 800-2000 MHz range during the
impulsive flare phase. Moreover, in some cases the spikes appear close to the
starting frequency of type III bursts (see Figure~\ref{fig1}b). It indicates
that the narrowband spikes in the 800-2000 MHz range are closely connected with
the primary flare energy-release process, i.e., with the flare magnetic
reconnection.

In papers by
\cite{1996SoPh..168..375K,2000SoPh..195..165K,2018SoPh..293..143K} it was
proposed that the spikes are generated in turbulence in the plasma reconnection
outflows. This suggestion was based on the Fourier analysis of the spike
frequency profiles. Therefore, let us make a similar analysis in the present
case. Considering that spikes are generated on Bernstein modes then the spikes
within the specific band of spikes correspond to one specific gyro-harmonic
number and frequencies $f$ of these spikes are proportional to the magnetic
field in their sources ($B \sim f$). We note that all bands of spikes in BM
model are generated in one region. Now, let us assume that the magnetic field
in this region changes as $B = B_0 \exp^{-z/H}$, where $B_0$ is the magnetic
field at some reference height in the solar atmosphere, $z$ is the height above
this reference level and $H$ is the height scale of the magnetic field. Then we
can re-sample the radio spectra expressed in frequencies to those expressed in
heights $z (f)$ according to the formula $z(f) = - H \ln \frac{B}{B_0} = - H
\ln \frac{f}{f_0}$, where $z(f)$ is the height in the solar atmosphere in
dependence on frequency $f$. For the purpose of the present study, we selected
the frequency band of 1370-1700 MHz that corresponds to the blue band at
12:26:15 UT in Figure~\ref{fig7} and also covers the following narrow band of
spikes (Figure~\ref{fig2}a). In order to increase the signal/noise ratio we
used the radio spectrum of this band with time resolution 0.2 s. Taking $f_0$ =
1700 MHz, we re-sampled this radio spectrum from f(MHz) to $z$. Then the
frequency profiles from these re-sampled radio spectrum were transform to the
Fourier spectra and these Fourier spectra were averaged over time intervals of
2 seconds. This interval of 2 s is taken as a compromise between highly
variable Fourier spectra at shorter intervals and only few spectra for longer
time intervals that do not show the Fourier spectrum evolution with
sufficiently good time resolution. As the result of these computations we
obtained the power-law spectra with the power-law index that evolved in time as
shown in Figure~\ref{fig8}. Examples of these spectra at three selected times
are shown in Figure~\ref{fig9}. Here, the interval of spatial scales in the
natural logarithm LN(k$_z$)= 4.2 -- 6.5, where the power-law index was
determined by the fitting techniques (black lines), is shown. As seen in
Figure~\ref{fig8} the power-law index evolves in time. Firstly, at the time of
strong broad band of spikes at 12:26:15 UT the Fourier spectrum
(Figure~\ref{fig9}) becomes very flat with the power-law index -0.8. Then the
power-law index decreases up to -2.4 at 12:26:32 UT. After this decrease, there
is the time interval of about 12:26:34 -- 12:26:45 UT with broad band of spikes
and the power-law index varying around the value -1.66 (-5/3). We note that
averaging of the radio spectra over this interval gives the power-law index
close to -5/3. In the following times the Fourier spectrum becomes steeper with
the power-law index up to -2.75. The final increase of the power-law index is
given by signal decrease in comparison to instrumental noise. The present
values of the power-law index are within the interval of the power-law indices
found in the paper by \cite{2000SoPh..195..165K} (-0.8 -- -2.85). It also
indicates that the uniqueness of this spike event is only in observations of
these very narrow bands of spikes. We note that in our previous study we
analyzed the whole 800-2000 MHz range, neglecting the bands of spikes and
considering different emission mechanism. On the contrary, here we have bands
of spikes that are interpreted as generated on Bernstein modes. Therefore, we
analyzed by this Fourier method only one band of spikes (1370-1700 MHz).
Nevertheless, we interpret the power-law spectra by the same way as in our
previous studies, i.e. the spikes are generated in the magnetohydrodynamic
turbulence, that in the stationary state is described by the Kolmogorov
spectral index -5/3. We think that the deviations of the found power-law
indices from the Kolmogorov spectral index is given by the evolution of this
turbulence and thus by deviations of this turbulence from the stationary
state.

Now, considering these results, we summarize the processes generating
the narrowband spikes in the 7 November 2013 event, as follows. In the flare
impulsive phase, during the turbulent magnetic reconnection, electrons are
accelerated as described in
\cite{2014PhPl...21i2304D,2015ApJ...806..167G,2016ApJ...827...94Z,2020PhPl...27h0501G}.
Turbulent reconnection and turbulent reconnection outflows consist many
magnetic traps of different sizes limited by magnetic mirrors. In these
magnetic traps the accelerated electrons form the loss-cone distribution
superimposed on the Maxwellian background plasma. As was shown by our
computations these electrons generate the Bernstein modes with the
gyro-harmonic numbers $s$ = 3, 4, 5 and 6. Then these electrostatic
Bernstein modes (BM) are transformed by their coalescence with the
low-frequency waves or by the coalescence of the counterpropagating Bernstein
modes into the electromagnetic waves as spikes on BM or double BM frequencies.

In turbulence that is in plasma reconnection outflows and/or in the
flare turbulent magnetic reconnection, the plasma parameters vary not only due
to the turbulence, but also due to gravitational stratification. Thus, in the
regions with this turbulence, where magnetic traps are formed, the ratio of the
electron plasma and electron-cyclotron frequencies $Y =
\omega_\mathrm{pe}/\omega_\mathrm{ce}$ varies in space and time. We think that
individual spikes are generated inside the magnetic traps. When a region with
the magnetic traps is small, i.e. with the very narrow interval of
values of $Y$, then only narrow bands of spikes are generated as in the ending
part of the 7 November 2013 event. On the other hand, for large regions with
such traps, i.e. with the broad interval of values of $Y$, broad bands of
spikes can be observed, see e.g. the broad bands of spikes in Figure~\ref{fig1}
in the 12:26:09 -- 12:26:39 UT interval. In turbulence the magnetic traps have
different sizes and thus individual spikes have different frequency bandwidths.
We think that this relation between sizes of the magnetic traps (in
correspondence with spatial scales in turbulence) and the frequency bandwidths
of spikes explains why the Fourier analysis of the frequency profiles of the
spike events show the power-law spectra.

\section{Conclusions}

We analyzed the narrowband spikes observed in the 800-2000 MHz range
during the impulsive phase of the November 7, 2013 flare. These spikes started
as typical spikes clustered in  broad bands and then the frequency distribution
of this spikes in the radio spectrum changed to four very narrow bands of
spikes. We focused our attention to these very narrow bands. We successfully
fitted frequencies of these bands by those calculating dispersion branches and
growth rates of the Bernstein modes. Using this model and taking the spike band
frequencies as BM frequencies we estimated the plasma density and magnetic
field strength in the narrowband spike source as $n_\mathrm{e}$ = 9.02 $\times$
10$^{9}$ cm$^{-3}$ and $B$ = 113 G, respectively. On the other hand, for the
radio emission on the double BM frequencies the values of the plasma density
and magnetic field strength are 4 and 2 times lower, respectively. We also
considered the UHW model \citep{1999ApJ...524..961S,2001A&A...379.1045B}. Comparing
the both models we presented arguments in favor of the BM model. Namely, we
succeeded (almost perfectly) to fit BM frequencies with the observed band
structure as expected in the BM model. Moreover, the frequency difference among
three bands was close to 300 MHz. It also speaks in favor of the BM model, because
it is not very probable that UHW model with three sources at different
locations give such a result. Based on these arguments, we concluded that the
BM model, proposed by \cite{1996ApJ...467..465W}, explains this spike event.

The  presented spike event shows a transition from typical spike event
with the broadband clouds of spikes to very narrow spike bands. It shows that
the BM model, that successfully fitted the narrow bands of spikes in the November
7, 2013 event, can be also valid at least for some group of spike events.
However, we think that UHW model or direct ECM models cannot be excluded in
some other spike events.

In the Fourier analysis of the frequency profiles of the November 7,
2013 event we found the power-law spectra with the power-law indices in the
-0.80 -- -2.75 interval. Because at some time interval this power-law index was
close to the Kolmogorov index (-5/3), we interpreted these power-law spectra in
a scenario where the narrowband spikes are generated in turbulent plasma
reconnection outflows or/and directly in the turbulent magnetic reconnection of
solar flares.

\acknowledgments M.K. and J.B acknowledge support from the project RVO-67985815
and GA \v{C}R grants 19-09489S, 20-09922J, 20-07908S and the financial support
by the German Science Foundation (DFG) via the projects BU-777-17-1. This work
was supported by The Ministry of Education, Youth and Sports from the Large
Infrastructures for Research, Experimental Development and Innovations project
``e-Infrastructure CZ LM2018140''. J.R. acknowledges support by the Science
Grant Agency project VEGA 2/0048/20 (Slovakia). Help of the Bilateral Mobility
Project SAV-18-01 of the SAS and CAS is acknowledged as well. We also
thank the anonymous referee whose comments helped to improve this paper.

\section*{Appendix A}
The dispersion branches of the electrostatic waves are computed using the
plasma permittivity tensor $\epsilon_\parallel$. We assume that plasma
consists of cold background Maxwellian electrons of density $n_\mathrm{e}$
and loss-cone hot electrons with density $n_\mathrm{h}$, where $n_\mathrm{e}
\gg n_\mathrm{h}$. Hot electrons have (DGH) distribution function for $j=1$
\citep{1965PhRvL..14..131D}
\begin{equation}
f_\mathrm{hot}(u_\parallel, u_\perp) = \frac{u_\perp^2}{2 (2\pi)^{3/2} v_\mathrm{t}^5} \exp \left(-\frac{u_\perp^2 + u_\parallel^2}{2 v_\mathrm{t}^2}\right),
\end{equation}
where $u_\parallel, u_\perp$ are the velocities parallel and perpendicular to
the magnetic field, respectively. $v_\mathrm{t}$ is the characteristic
velocity. All ions have same temperature as cold electrons. In our case,
ions do not significantly contribute to the permittivity for frequencies
$\omega \sim \omega_\mathrm{pe}$, where $\omega_\mathrm{pe}$ is the electron
plasma frequency.

For condition $n_\mathrm{h} \ll n_\mathrm{e}$, the permittivity can be
separated into the permittivity $\epsilon_\parallel^{(0)}$ for cold
electrons and the permittivity $\epsilon_\parallel^{(1)}$ connected with hot
electrons \citep{1974itpp.book.....C,1997riap.book.....Z,fitzpatrick}. In this
approximation, solutions for the electrostatic waves are given only by the term
$\epsilon_\parallel^{(0)} = 0$. The relation for dispersion branches of
the Bernstein modes can be written in the form
\begin{equation}
\epsilon_\parallel^{(0)} = 1 - 2\omega_\mathrm{pe}^2 \frac{e^{-\lambda}}{\lambda} \sum_{l=1}^{\infty} \frac{l^2 I_l(\lambda)}{\omega^2 - l^2 \omega_\mathrm{ce}^2 } = 0,
\label{eq-root}
\end{equation}
\begin{equation}
\omega_\mathrm{pe}^2 = \frac{n_\mathrm{e} e^2}{m_\mathrm{e}
\epsilon_\mathrm{0}}, \qquad \lambda = \frac{k_\perp^2
v_\mathrm{tb}^2}{\omega_\mathrm{ce}^2},
\end{equation}
where $\epsilon_\mathrm{0}$ is the permittivity of free space,
$\omega_\mathrm{ce}$ is the the electron cyclotron frequency, $\mathbf{k} =
(k_\parallel, k_\perp)$ is the wave vector parallel and perpendicular to the
direction of the magnetic field, respectively, $\omega$ is the wave frequency
of the electrostatic wave, $I_l(\lambda)$ is the modified Bessel function of
$l$th order, $\lambda$ is the dimensionless parameter, $m_\mathrm{e}$ is the
electron mass, $e$ is the electron charge.

We searched roots (dispersion branches) directly from Equation~\ref{eq-root}
using Python and Scipy library\footnote{python.org, scipy.org}; we do not use
any approximative analytical solutions. Specifically, we search for roots using
the Levenberg--Marquardt damped root method
\citep{Levenberg,Marquardt,More:126569,Press:2007:NRE:1403886} and following
the method by \cite{2019ApJ...881...21B}. Our method divides the examined
$\omega - k_\perp$ domain into a rectangular grid. Each grid cell then serves
as a starting point in the root search algorithm. We used $10^3$ starting
points in $\omega$ and $400$ starting points in $k_\perp$ direction. Because
the branches can be usually assumed as horizontal, the minimalization is made
in variable $\omega$. The minimalization procedure has the eventual error $ <
10^{-8} \omega_\mathrm{pe}$. We found that for such an error, sum over harmonic
number $l \leq 30$  is sufficient. Consequently, all the found roots from all
starting points are aggregated, giving the resulting dispersion branches.

Generally, the electrostatic (Bernstein and upper-hybrid) waves can be unstable if
the double plasma resonance condition is fulfilled
\begin{equation}
\omega - \frac{k_\parallel u_\parallel}{\gamma_\mathrm{rel}} - \frac{s \omega_\mathrm{ce}}{\gamma_\mathrm{rel}} = 0,
\end{equation}
where $\gamma_\mathrm{rel} = (1 - v^2 / c^2)^{-\frac{1}{2}}$ is the
relativistic Lorentz factor and $s$ is the resonance gyro-harmonic number.
Whether amplitude of an electrostatic wave increases for some
$(\omega,k_\perp)$ we need to calculate the growth rate
\citep{1975SoPh...43..431Z}
\begin{equation}
\gamma(\omega, k_\perp) = - \frac{ \mathrm{Im} \, \epsilon_\parallel^{(1)}}
{\left[\frac{\partial \mathrm{Re} \, \epsilon_\parallel^{(0)}}
{\partial \omega}\right]_{\epsilon_\parallel^{(0)} = 0}},
\end{equation}
\begin{equation}
\frac{\partial \epsilon_\parallel}{\partial \omega} = 4 \omega \omega_\mathrm{pe}^2 \frac{e^{-\lambda}}{\lambda} \sum_{l=1}^\infty \frac{l^2 I_l(\lambda)}{(\omega^2 - l^2 \omega_\mathrm{ce}^2)^2}.
\label{eq10}
\end{equation}

\begin{figure}
\begin{center}
\includegraphics[width=10cm]{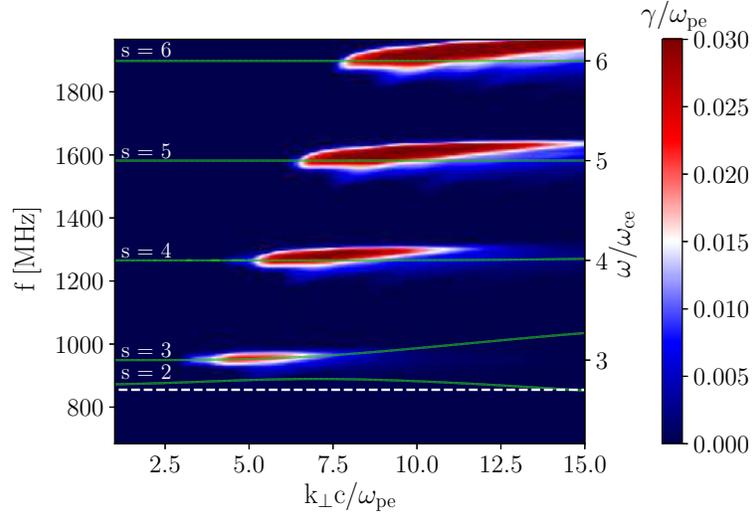}
\end{center}
  \caption{Growth rates of the Bernstein modes as a function of the frequency
  and perpendicular k-wave vector for parameters $\omega_\mathrm{pe}/\omega_\mathrm{ce}=2.7,
  v_\mathrm{tb}/c=0.02, v_\mathrm{t}/c=0.25, n_\mathrm{e}/n_\mathrm{h} = 10$.
  The values of f in MHz correspond to the radio emission on BM frequencies, i.e. $\omega_\mathrm{pe}= 2 \pi f_\mathrm{pe}$, $f_\mathrm{pe}
= 855$~MHz. For the radio emission on double BM frequencies the values of f need be divided by 2 and $f_\mathrm{pe}$ = 427.5 MHz.
   \textit{Green lines:}
  Dispersion branches computed from the Equation~\ref{eq-root}, $s$ is the gyro-harmonic number of each branch.
  \textit{White dashed horizontal line:} The plasma frequency.}
  \label{fig5}
\end{figure}

The explicit expression for this growth rate was provided by
\citet{2005A&A...438..341K}. The imaginary part of the permittivity can be
written as follows
\begin{multline}
\mathrm{Im}(\epsilon_\parallel^{(1)}) = - 2\pi^2 m_\mathrm{e}^4  \frac{\omega_\mathrm{pe}^2}{k^2}
    \sum_{l=s+1}^{\infty} ab^2 \\
    \times \int_0^{\pi}  J_l \left( \frac{\gamma_\mathrm{rel} k_\perp v_\perp}{\omega_\mathrm{ce}} \right)
    \frac{\gamma_\mathrm{rel}^5 \sin \phi}{\frac{\partial \psi}{\partial\rho}}  \frac{l\omega_\mathrm{ce}}{\gamma_\mathrm{rel} v_\perp}\frac{\partial f}{\partial p_\perp} \, \mathrm{d}\phi,
\label{eq-growth}
\end{multline}

\begin{equation}
\frac{\partial \psi}{\partial \rho} = \frac{\gamma_\mathrm{rel}^2 l \omega_\mathrm{ce}}{c^2} \left( v_\parallel^2 + v_\perp^2 \right),
\end{equation}

\begin{equation}
    v_\parallel = - a \cos(\phi), \quad v_\perp = b \sin(\phi),
\end{equation}

\begin{equation}
    a^2 = \frac{l^2 \omega_\mathrm{ce}^2 c^2 (l^2 \omega_\mathrm{ce}^2 - \omega^2)}
    {l^4 \omega_\mathrm{ce}^4}.
\end{equation}
As positions of the growth rates in Figure~\ref{fig5} are given only by hot
electrons, we can compute the growth rates independently on position of
dispersion branches in the whole domain $\omega-k_\perp$ using
Equation~\ref{eq-growth}. We use 600 grid points in $\omega$ direction, and
$300$ points in $k_\perp$ direction. The growth rates are overlaid by solutions
of the dispersion equation (Equation~\ref{eq-root}) that represent the
Bernstein modes used in the interpretation of spikes in the model by
\cite{1996ApJ...467..465W}.

\begin{figure}
\begin{center}
\includegraphics[width=8cm]{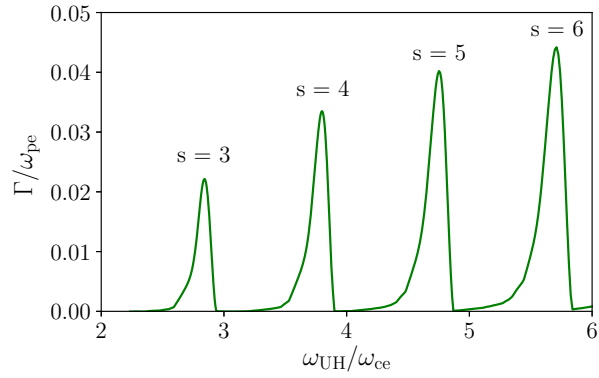}
\end{center}
  \caption{Growth rate of the upper-hybrid waves in dependence on the $\omega_\mathrm{UH}/\omega_{ce}$ ratio,
  $s$ denotes the gyro-harmonic number.}
  \label{fig6}
\end{figure}

\begin{figure}
\begin{center}
\includegraphics[width=8cm]{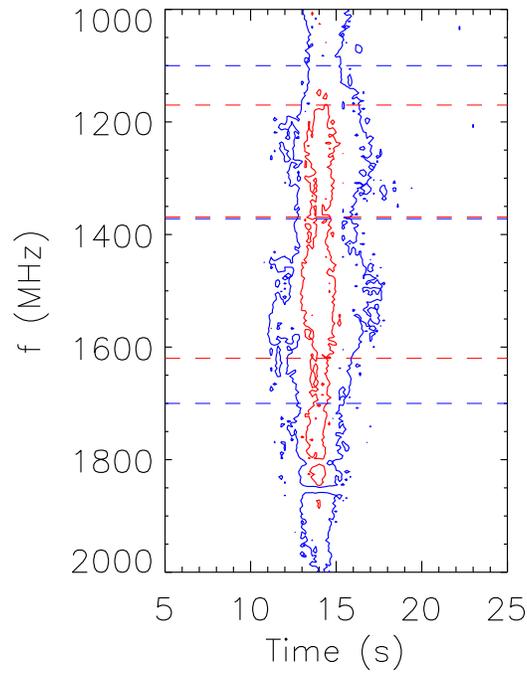}
\end{center}
  \caption{Blue and red contours of spikes at two levels (320 a.u. and 440 a.u.) at 12:26:05-12:26:25 UT
  (Time (s) after 12:26:00 UT), compare with spikes in Figure~\ref{fig2}a.
    Blue and red horizontal dashed lines limit the frequency
  width of two bands of spikes as 1100 - 1370 MHz and 1370 - 1700 MHz, and as 1170 - 1370 MHz and 1370 - 1620
MHz, respectively.}
  \label{fig7}
\end{figure}

\begin{figure}
\begin{center}
\includegraphics[width=10cm]{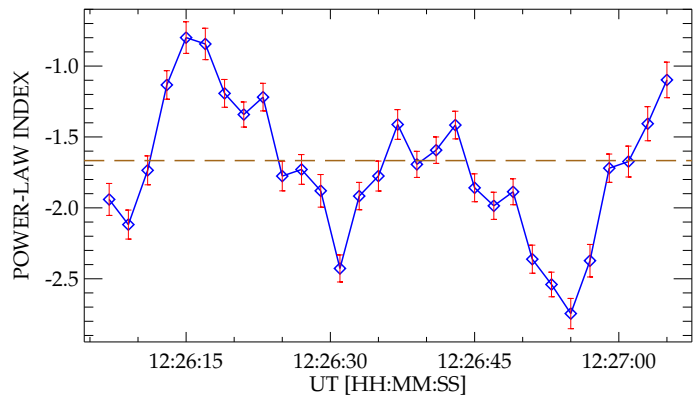}
\end{center}
  \caption{Time evolution of the power-law index of the Fourier spectra of the 1370-1700 MHz frequency profiles during the November 7, 2013 event.
  The horizontal dashed line corresponds to the power-law index -5/3.}
  \label{fig8}
\end{figure}

\begin{figure}
\begin{center}
\includegraphics[width=10cm]{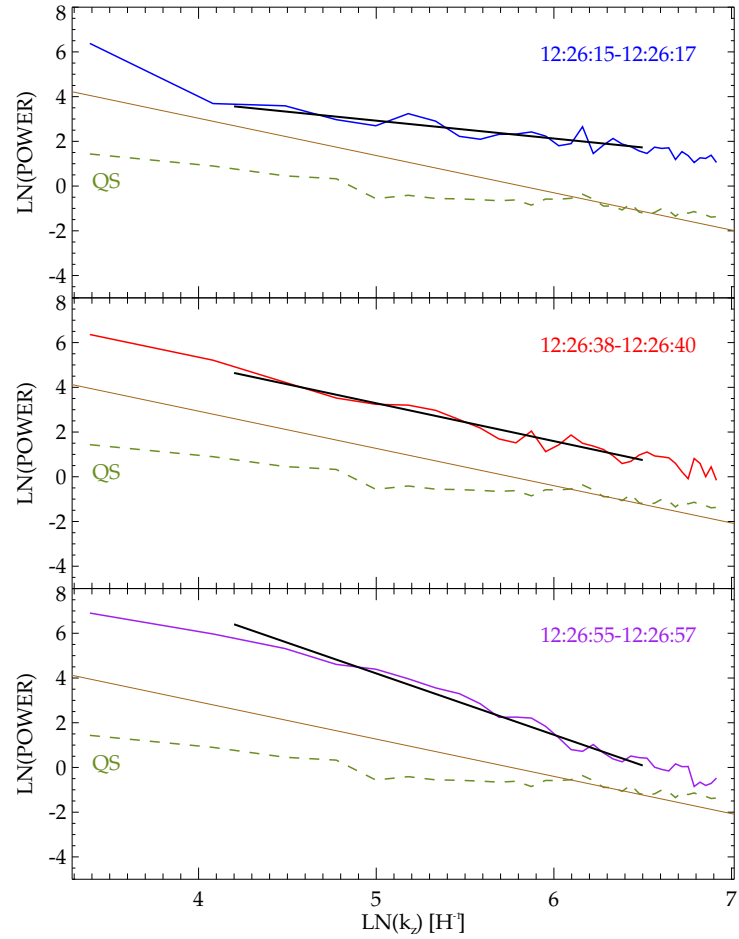}
\end{center}
  \caption{Fourier spectra of the 1370-1700 MHz frequency profiles integrated over 2 s at three instants 12:26:15 UT (blue line),
  12:26:38 UT (red line) and 12:26:55 UT (violet line), in the natural logarithmic scales (LN)
  together with the
  spectrum of the frequency profile without the emission before the spike event (QS - quiet Sun) at 12:26:00 UT (olive dashed line).
  Black straight lines mean the fitted lines of the Fourier spectrum in the interval of LN(k$_z$)= 4.2 -- 6.5. The olive straight lines denote the lines
  with the power-law index -5/3.}
  \label{fig9}
\end{figure}

On the other hand, in the interpretation of spikes generated at the
upper-hybrid frequency, we compute the growth rates in Figure~\ref{fig6} as
follows: First, for a given ratio $\omega_\mathrm{UH} / \omega_\mathrm{ce}$ a
branch of the upper-hybrid waves is calculated from the relation
$\omega_\mathrm{UH} = \sqrt{\omega_\mathrm{pe}^2 + \omega_\mathrm{ce}^2 +
3k_\perp^2 v_\mathrm{tb}^2}$. Then for this branch the growth rates from
Equation~\ref{eq-growth} are computed and among them the maximal growth rate
\begin{equation}
\Gamma \left( \frac{\omega_\mathrm{UH}}{\omega_\mathrm{ce}} \right) = \textrm{max} \left\{ \gamma_\mathrm{UH} \left( \frac{\omega_\mathrm{UH}}{\omega_\mathrm{ce}}, \omega, k_\perp \right) \right\},
\end{equation}
is taken as a value for the given ratio $\omega_\mathrm{UH} /
\omega_\mathrm{ce}$.


\newpage

\end{document}